\documentclass[10pt,conference,USletter]{IEEEtran}
\IEEEoverridecommandlockouts
\usepackage{cite}
\usepackage{amsmath,amssymb,amsfonts}
\usepackage{algorithmic}
\usepackage{fancyhdr}
\usepackage{graphicx}
\usepackage{textcomp}
\usepackage{xcolor}
\usepackage{hyperref}
\usepackage[singlespacing]{setspace} 
\def\BibTeX{{\rm B\kern-.05em{\sc i\kern-.025em b}\kern-.08em
    T\kern-.1667em\lower.7ex\hbox{E}\kern-.125emX}}
\begin{document}

\title{A Survey of Mobile Edge Computing for the Metaverse: Architectures, Applications, and Challenges}

\author{\IEEEauthorblockN{Yitong Wang\\yitong.wang@ntu.edu.sg}
\IEEEauthorblockA{\textit{School of Computer Science and Engineering} \\
\textit{Nanyang Technological University}\\
\textit{Singapore} \\
}
\and
\IEEEauthorblockN{Jun Zhao\\ junzhao@ntu.edu.sg}
\IEEEauthorblockA{\textit{School of Computer Science and Engineering} \\
\textit{Nanyang Technological University}\\
\textit{Singapore} 
}
}

\maketitle
 \thispagestyle{fancy}
\pagestyle{fancy}
\lhead{This paper appears in  the 2022 IEEE 8th International Conference on Collaboration and Internet Computing (CIC).\\ Please feel free to contact us for questions or remarks.}

\begin{abstract}
Metaverse is an emerging virtual universe where humans can have real-time interactions and solid social links like in the physical world, and it opens up a new era of Internet and interactions. In Metaverse, an immersive and photorealistic environment promotes social activities, including education, meetings, and shopping of digital avatars based on critical technologies, including 3D rendering, extended reality, digital twins, artificial intelligence, and Blockchain. However, the limitations of computation, storage, and energy resources restrict the development of Metaverse, and a series of system issues (e.g., latency, security, and battery-life)  continue to arise. As a result, how to find corresponding measurements to mitigate unsatisfactory influences becomes the focus. Mobile edge computing (MEC) as a distributed computing paradigm offloads computation-intensive tasks to the edge of the network. It brings the resources as close as possible to the end devices, addressing the shortcomings mentioned above. In this paper, we propose a comprehensive survey of the MEC-based Metaverse. Particular emphasis is given to the technologies convergence, architectures, and application scenarios, e.g., BoundlessXR and CloudXR. Significantly, we introduce the potential future directions for developing Metaverse systems.
\end{abstract}


\section{INTRODUCTION}
Metaverse, as a universe of ternary-world interactions  (i.e., physical world, human world, and digital world), enables humans as digital avatars in sub-Metaverses to carry out various social activities in the physical world, such as business meetings,  learning, shopping, and digital assets trading\cite{zhong2022empowering}. Metaverse significantly revolutionises human manners of interactions with machines, from conventional 2D web browsing to immersive real-time interaction in 3D virtual environments, supported by high fidelity image technology, 3D rendering, Blockchain, digital twins, artificial intelligence (AI) and sophisticated sensor devices. Recently, Metaverse has been investigated by lots of commercial and gaming companies. For example, Tencent and Facebook have already invested in chat scenarios of Metaverse; even Facebook renamed itself ``Meta" in 2021.

The increasing number of mobile devices and users has caused the fact that wireless communication networks constantly evolved to cope with the strict requirement of high data rate, latency, and machine-to-machine(M2M) connection density\cite{popovski20185g}. 5th generational communication network that consists of specific attributes is the best choice to cater for the emerging Metaverse system. The advent of 5G boosts the performance indicators of communication networks by a multiple of 10 to 100 times, which meets the stringent requirements of various services and applications. It acts as a bridge for the transmission of original pose data and compressed image data generated by VR devices and edge cloud servers, respectively, which perfectly enhances the rendering performance of the Metaverse.

Extended reality (XR)\cite{guan2022extended} technology incorporates mixed reality (MR), augmented reality (AR), virtual reality (VR), and any other technology that encompasses the fusion of all reality and virtualization. XR devices collect data, including biological data like eye-, hand-, and head-tracking, and the accumulation of user data from other social media platforms\cite{evans2022white}. XR applications also provide multi-sensor immersiveness and real-time interactions for users. Augmented reality(AR)\cite{torres2022conceptualization}, as a technology that replicates the sensory perception of the real world in terms of time and space domains, projects additional augmentations generated by computers upon natural objects, combining real and virtual worlds to enhance people's sensory experience further. In the last few years, both KLM Royal Dutch Airlines and British budget airline EasyJet have allowed passengers to check the size of their suitcases for boarding through AR technology. AR is performed based on realistic environments, and application services enrich reality with augmented components\cite{braud2022scaling}. On the opposite side, VR is a part of MR where its surrounding environment is virtual, and people can fully have immersive experiences in virtual worlds with appropriate wearable equipment (e.g., VR Helmets or Glasses)\cite{qu2022digital}. To sum up, VR provides immersive interactivity in a digital world. AR delivers authentic experiences to customers with digital holograms, images, and videos based on real-world objects. MR acts as a transition, providing a distinctive experience between VR and AR\cite{wang2022survey}. Further developments in micro-sensors and XR technology are making XR equipment, such as helmet-mounted displays (HMDs), promising to be the primary endpoint for experiencing the Metaverse~\cite{sugimoto2021extended}.

The exponential growth of mobile internet traffic has driven a dramatic evolution in computing paradigms. Modern mobile applications have more stringent requirements which cloud computing cannot satisfy, such as ultra-low latency, ultra-high throughput, ultra-high stability, and high spectral efficiency. Meanwhile, big data-associated Internet of Things devices causes an immense growth in the total traffic flows. To break the limitations of cloud computing and ensure better customer service performance, the emerging computing paradigm of edge computing, especially mobile edge computing, has gained the spotlight in both commercial and academic fields nowadays.

Following European Telecommunications Standard Institute (ETSI), Mobile Edge Computing (MEC) is formally defined as a new platform with significant computing resources, including IT and Cloud-computing and it gets closer to the subscribers' side \cite{computing2014mobile}. Intending to extend 3GPP access scenarios such as Wi-Fi, the original MEC concept has now transitioned to ``Multi-access Edge Computing" by ETSI. The remarkable characteristic of MEC is offloading computation tasks from core data centres to distributed edge servers located on the base stations. Thus, this typically distributed computing paradigm significantly reduces the communication latency and provides a relatively large amount of computation resources compared with conventional cloud computing paradigms. Also, the deployment of MEC relies heavily on the virtualized platforms\cite{hu2015mobile}. 

Based on the evolution and combination of the various technologies, services, and equipment mentioned above, Metaverse technology is rapidly gaining attention and enhancement. However, there is a highly distant path to explore to realise the Metaverse fully. Firstly, transmitting the surging data flows together to cloud servers from end devices is contrary to what is being sought. The reason is that it not only aggravates network latency on the system but also causes network congestion and data loss when reaching cloud services via the core network. Furthermore, data leakage and asset authentication of the Metaverse are both obstacles encountered. Data security leads to the loss of all personal information and the forgery of avatar identities. As multiple administrators provide the assets and virtual currencies in Metaverse, how to authenticate and trade among the diverse assets in sub-Metaverses raises a concern.

The contributions of our survey are as follows:
\begin{itemize}
\item We first introduce the relevant background and basic concepts of Metaverse and present a clear overview of the types of architectures, evolutionary and eventual outcomes in Metaverse. Notably, we provide a nuanced introduction to indicators and essential technologies of Metaverse. With these concise introductions, readers can keep abreast of the ongoing technological developments in Metaverse.
\item We discuss the technological convergence of MEC and Metaverse and illustrate the role that MEC plays in Metaverse. Unlike previous papers on MEC-based Metaverse, we not only describe the contributions of MEC to various performance indicators but also emphasize the application of collaborative architectures of MEC with other computing paradigms in Metaverse. Specially, we point out that MEC is crucial in solving the latency, privacy, and energy challenges in Metaverse.
\item We outline the research directions and challenges to pave the path toward future research attempts to realize higher performance in MEC-based Metaverse architectures. Furthermore, our survey aims to assist researchers in gaining a thorough and in-depth understanding of the MEC-based Metaverse and grasping a holistic view of the research in this area.
\end{itemize}
The remainder of this paper is organized as follows. Section~\ref{sectionII} introduces the present advanced developments of technologies in Metaverse, and the scenarios of mobile edge computing applied to the XR domain. Section~\ref{sectionIII} interprets the concepts, architectures, and features of the Metaverse and identifies some of the current challenges. An introduction to mobile edge computing, including its strengths and weaknesses, is presented in Section~\ref{sectionIV}. Section~\ref{sectionV} summarises the enhancements and facilitations when mobile edge computing is applied to the Metaverse and demonstrates the architectures of the Metaverse based on the collaboration of mobile edge computing with cloud and fog computing. Future research directions are shown in Section~\ref{sectionVI}. Section~\ref{sectionVII} provides the conclusion of the paper.
\begin{figure*}
    \centering
    \includegraphics[width=0.8\textwidth, height= 0.5\textwidth]{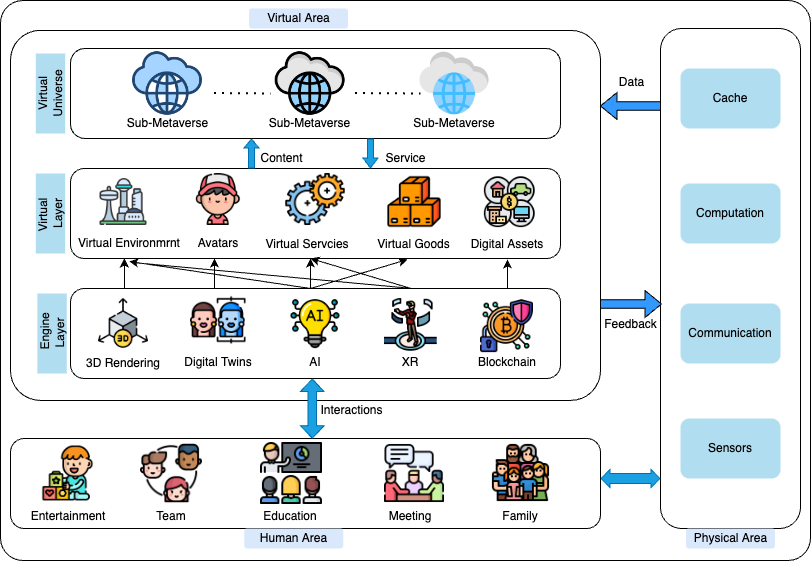}
    \caption{The Metaverse architecture performs the interactions among virtual area, human area, and physical area}
    \label{fig1}
\end{figure*}

\section{RELATED WORK}\label{sectionII}
We have witnessed many papers in terms of different aspects of Metaverse technologies and application scenarios of mobile edge computing. In this section, we discuss recent Metaverse and MEC research and show the main distinction with our paper. \textit{Fernandez et al.}\cite{fernandez2022life} review an in-depth survey of Metaverse from three main motivation perspectives, including privacy, governance, and ethics design. Specifically, they analyze privacy from sensory level, behaviour and communications, as well as users and bystanders. Also, further challenges of these three factors and a novel modular-based framework are presented in this paper. \textit{Xu et al.}\cite{xu2022metaverse} explore the significance of using specific spectrums in Metaverse and investigate the vision of Blockchain technology, respectively. \textit{Cheng et al.}\cite{cheng2022will} focus on some existing social VR platforms that have the potential to evolve into the architectures of the Metaverse and also conduct a test by looking in depth at the network operation and capabilities of two typical platforms. \textit{Lee et al.}\cite{lee2021creators} present several innovations in the computation arts of Metaverse, proposing some research agendas about democratizing computational arts, digital privacy and safety for Metaverse artists, the identification of proprietary rights in digital artworks, and the directions of technological developments. For MEC, \textit{Siriwardhana et al.}\cite{siriwardhana2021survey} emphasize the survey of mobile augmented reality (MAR) based on MEC and discusses future vital application areas for MAR. \textit{Lim et al.}\cite{lim2022realizing} introduce a study focused on edge-intelligence Metaverse in term of virtual city development. A stochastic optimal resource allocation scheme (SORAS) is further proposed by\textit{Ng et al.} based on stochastic integer programming with the aim of optimizing the cost-effectiveness of virtual service suppliers \cite{ng2021unified}. However, none of these recent papers focuses on the convergence of the Metaverse and MEC paradigm in terms of unique perspectives such as quality of experience, application scenarios, future challenges, and so on.

\section{METAVERSE}\label{sectionIII}
\subsection{Metaverse Introduction}
The term 'Metaverse' first emerged in the 20th century in Neal Stephenson's science fiction novel named 'Snow Crash'\cite{di2021metaverse}, but did not receive much attention at that time. Two primary factors have driven the Metaverse into focus currently: 

\begin{itemize}
\item \textit{Social factors}: The Covid-19 pandemic has led to a dramatic shift in the manners of people living and working via the Internet from outdoors to indoors\cite{bates2021global}, which has caused an increased frequency of daily activities via the Internet. Additionally, commercial enterprises' strong focus on VR resulted in the re-emergence of the Metaverse in the public arena;
\end{itemize}

\begin{itemize}
\item \textit{Technical factors}: XR ecosystem developments (including AR, VR, MR) facilitate the further enhancement of virtual scenarios and interactions between physical and virtual worlds; Emerging Web 3.0 converts ``server-centric" networks into ``user-centric" networks based on decentralized networks; MEC servers perform data processing at the edge of the network (i.e., closer to the user) to mitigate system latency; 5G networks with downlink (DL) speeds of 200Mbps meet the constraints including low latency, high throughput, and reliability of the Metaverse. 
\end{itemize}

Through the ternary nature of the Metaverse definition, Fig.~\ref{fig1} shows the architecture of the Metaverse, and the relationships between these three worlds can be explained in detail as follows:  

1) \textit{Physical Area:} The physical area provides the essential infrastructures to support the other two worlds, including computation, cache, transmission, and sensors. Data from around the environment and the human body are collected precisely by sensor infrastructures. Transmission infrastructures (e.g. artificial satellites strictly) ensure the stability and continuity of the entire network connection. Computationally intensive tasks are processed by servers located at different positions via robust computation infrastructures, depending on the type of these tasks. Also, cache infrastructures at various locations (i.e., local cache, edge cache, and cloud cache) can store major tasks and reduce the total latency of the entire task processing.

2) \textit{Virtual Area:} In the Virtual world, the Metaverse engine facilitates and manages large volumes of digital information from both the physical and human worlds to enable large-scale Metaverse services\cite{wang2022survey}. 
\begin{itemize}
\item \textit{Engine Layer:} Metaverse engine layer consists of various technologies to provide a photorealistic environment, real-time interactions, non-fungible token (NFT)\cite{garcia2022semantics} and translation, respectively. 3D rendering technology provides a virtual environment with diverse spatiotemporal dimensions and attributes. Digital twins (DT) is crucial to the behaviours of avatars in Metaverse. Especially various human social activities are identified and performed via DT technology. AI plays a vital role in guaranteeing the reliability of Metaverse. By combining with other technologies, many machine learning algorithms are used to solve challenging tasks, such as computational resource allocation\cite{si2022resource}, predictions of human actions, and increased spectrum utilisation.
\item \textit{Virtual Layer:} As illustrated in Fig.~\ref{fig1}, the virtual layer is composed of the virtual environment, digital avatars, virtual services/goods and digital assets, which is supported by the critical technologies in the engine layer. High-fidelity images are provided with 3D rendering and AI support by collecting and processing real-world environmental data via XR sensor devices. Digital avatars are capable of behaving exactly like humans in the real world based on DT technology.
\item \textit{Virtual Universe:} In contrast to the concept of a unified Metaverse, the whole Metaverse comprises a series of dispersed sub-Metaverses where avatars can be served distinctly. Distinct sub-Metaverses can be freed from the constraints of time and space. Due to the existence of multiple operators, creating multiple sub-universes becomes a fact, but their interconnection holds the key to future developments.
\end{itemize}

3) \textit{Human Area:} In reality, people’s intricate social relationships and diverse activities contribute to the functioning of society. Crucially, as in the real world, people as avatars are at the centre of the virtual world and are constantly creating virtual objects. Currently, by wearing devices (e.g. XR HMDs, Wristband sensors), social activities such as business meetings, office learning and entertainment concerts can be realised in the virtual world via human-computer interaction (HCI) and especially brain-computer interaction (BCI). Human activities can be accurately mapped onto the virtual world through various Metaverse engines. And services in the virtual world can also continuously influence human interactions and perceptions in the real world. 

\subsection{Metaverse Characteristics}
\textit{Immersive:} The ultimate goal of Metaverse development is to approximate the user's experience in the real world gradually. Supported by various infrastructures and Metaverse engines, users can interact in a rendered and high-fidelity virtual universe. Immersive means that the user can adapt to the digital world and move around in the Metaverse without discomforts, such as dizziness and nausea.

\textit{Interoperability:} Multiple operating systems, such as Ethereum, Solana, and Polygon, co-exist in the Metaverse. Accordingly, a variety of value tokens that are derived from each technology continuously emerge. For example, the Bitcoin Blockchain supports a token called Bitcoin, while the Ethernet-backed tokens include SAND, MANA, AXS, and GALA. The interoperability of the Metaverse is guaranteed by the accessibility of connections and transactions between different sub-universes or currencies.

\textit{Multi-technology:} A few technological advancements have contributed to the emergence of the Metaverse. For example, 5G technology has improved peak transmission rate, time delay, and reliability to meet the requirement that transmissions must be completed in milliseconds or less, relieving users of dizziness caused by high latency (\textgreater20ms). Also, because downlink speed is proportional to the rendering resolution, higher downlink speeds can make virtual environments more realistic. In summary, multi-technology provides an immersive experience and high-fidelity virtual worlds based on augmented reality technology, generates a city twin of the physical society with digital twin technology \cite{han2022dynamic}, and creates an economic system based on Blockchain technology \cite{ning2021survey}.

\subsection{Metaverse Technologies}
\textit{XR:} XR technology covers AR, VR, and MR. Wearable VR devices provide users with a completely virtual scenario in which they can be fully immersive\cite{xu2022wireless}. For example, Second life provides a virtual world in which players can survive. In this world, players can control their avatars to perform activities and enjoy personalized services. For AR/MR technology, augmentations are overlaid in the real world\cite{mcgill2021ieee}, e.g., users can play Pokémon GO on their mobile phones and catch virtual Pokemon. XR devices significantly enhance the user's immersion in the Metaverse.

\textit{Digital Twin:} Digital twin (DT) projects humans, objects, and environments of the real world into the virtual world, creating digital clones that are visually indistinguishable from the real world in real-time\cite{lynch2022smart}. By processing the input data, DT is able to manage and optimize physical objects periodically \cite{aloqaily2022integrating}\cite{van2022edge}. Significantly, the data flows between virtual objects and the physical world are bidirectional. The physical object transmits the collected data in multiple formats to the virtual twins, and the virtual twins convey the processed feedback to the physical object\cite{wu2021digital}, further accelerating the intersection of the human and virtual worlds.

\textit{Blockchain:} Blockchain is considered to be a decentralized technology \cite{wang2022blockchain}. Unlike traditional centralized systems, Blockchain technology can help safeguard users' data, identity information, and virtual assets with certificates (i.e., NFTs) from leakage and theft in the event of a threat. At the same time, Blockchain, as a distributed ledger, ensures that all data is protected from tampering and modification. Hence, Blockchain not only ensures data security and quality but also enables seamless data sharing and data interoperability and integrity\cite{gadekallu2022blockchain}. Furthermore, the convergence of Blockchain and other key technologies can significantly empower the performence of the Metaverse system \cite{badruddoja2022trusted}.

\textit{AI:} AI plays an indispensable role in creating and rendering large-scale Metaverse. Conventional AI technologies are composed of supervised learning, semi-supervised learning, unsupervised learning, as well as reinforcement learning (RL)\cite{lin2022icdvae}. Nowadays, many advanced machine learning (ML) algorithms of supervised and reinforcement learning have been used for different challenging tasks, including automatic resource allocation, attack prevention, and network fault detection\cite{huynh2022artificial}\cite{zhou2022resource}. Based on massive Metaverse multimodal input data, high-quality Metaverse scenes can be created and rendered by AI via considerable data interference\cite{bouachir2022ai}. Also, deep learning (DL) and ML algorithms facilitate the provision of personalized services by making good decisions. 

\section{MOBILE EDGE COMPUTING}\label{sectionIV}
\subsection{MEC Introduction}
\begin{figure}
    \centering
    \includegraphics[width=7.5cm, height= 8cm]{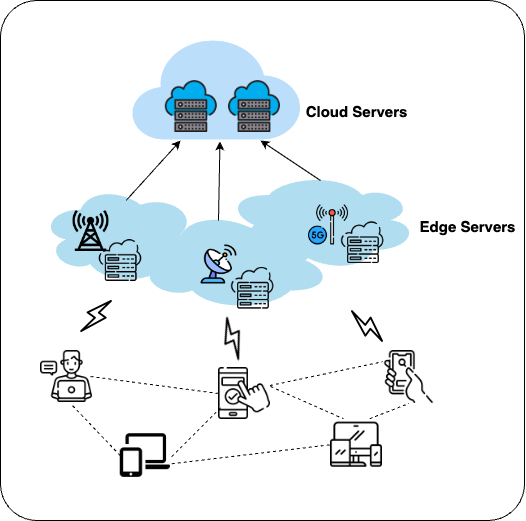}
    \caption{MEC computing paradigm features the low-latency and energy-efficiency procession.}
    \label{fig2}
\end{figure}

The motivation for investigating MEC computing paradigms is tackling the tradeoff between computation and communication. As shown in Fig.~\ref{fig2}, edge servers are deployed on the edge of networks, which provides computational resources. As a result, computational tasks are processed at the network's edge as proximate to the user as possible. Different from fog computing and cloud computing, a MEC server is a node device which means that the decentralization of mobile communication networks is achieved\cite{vhora2020comprehensive}. The security and privacy of networks are enhanced greatly, and congestion during peak transmission could also be tackled.

Nowadays, an increasing number of papers on mobile edge computing are investigated, especially in the fields of system architecture design and task offloading. Paper\cite{sheng2019computation} proposes a task offloading strategy based on an improved auction algorithm, divided into two main phases: task offloading and task scheduling, considering the limited resources of smart devices and the stronger computing and storage capabilities of edge servers. Furthermore, this strategy provides the basis for offloading decisions in the decision phase by considering the time cost of task execution locally or at the edge as well as energy consumption to determine whether the computation task that needs to be processed will be offloaded to the edge of the network. Once the task is offloaded to the edge server, reducing computational latency and reducing transmission energy consumption to achieve the global optimum are the main objectives of the task scheduling phase. Paper \cite{wang2021research} considers mobility at the user end, classifies cellular networks in more detail and gives more accurate state transfer probabilities. And this research models the VM migration process as a Markov decision process and uses a policy iteration algorithm to find the optimal solution, effectively reducing energy consumption and latency.

\subsection{MEC Application}
MEC is used as a key technology that guarantees the efficient realisation of diverse services, especially in the practical application of task offloading and resource allocation. In vehicular edge computing (VEC) networks\cite{liu2019deep}, vehicles can act as an edge server at the edge of the Internet of Things (IoT) system to provide computing resources\cite{li2022internet} and various services. Meanwhile, the collaboration between vehicle edge servers (VES) and fixed edge servers (FES) can provide a variety of offloads and computing options for network computing, offering various edge computing resources depending on the tasks. \textit{Ren et al.} \cite{ren2018mobile} demonstrate that edge servers can reduce the core network use by elaborating on the system framework of web-based services with MEC technology and further showing that MEC can address real-world Web AR deployments. 

\section{METAVERSE WITH MEC}\label{sectionV}
It is theoretically feasible to perform all computing tasks on the local XR devices or cloud servers, but this contradicts the desire to prolong battery life, lighten device weight and maintain ultra-low latency. The convergence of mobile edge computing and Metaverse results in a new generation of MEC technology\cite{zhang2017towards}\cite{dhelim2022edge}, which is being used to break the dependency of devices on centralized cloud servers and to improve the high real-time performance and security of the entire system. Primarily, splitting rendering is generated after this convergence. As graphics are heavily rendered, on-device processes are augmented by being partitioned to IoT devices and edge clouds. The graphical rendering of the Metaverse system is enhanced by using photonic processing for head tracking and motion tracking of latency-sensitive devices. Furthermore, based on the design of stream processing technology, Metaverse systems can also effectively decrease the latency and have excellent performance as a high-performance distributed stream processing system (DSPS)\cite{heo2021poster}. More studies are explained as follows.
\begin{figure}
    \centering
    \includegraphics[width=7.5cm, height= 5cm]{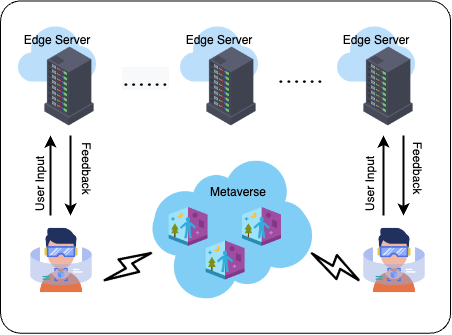}
    \caption{Architecture of Mobile Edge Computing-Based Metaverse}
    \label{fig4}
\end{figure}
\subsection{MEC-Based Architecture}
Conventional cloud-based Metaverse architecture is not beneficial to implementing virtual functions and services via transmitting massive data such as movements of avatars and physical factors of surroundings to the centralized server. To mitigate negative influences caused by limitations, innovative MEC-based Metaverse architectures have resulted in optimization and a satisfactory performance:

\textit{1) Mobile Edge Computing-Based Metaverse:} In the Metaverse of this architecture, multiple dynamic edge nodes are incorporated to perform users' instructions, which significantly decrease the delay caused by users' movements\cite{chang20226g}. As shown in Fig.~\ref{fig4}, the private data generated by the customers is transmitted to the specific edge server. Through the processing and delivery of the edge servers, customers enjoy a low-latency service and an immersive Metaverse experience. To cope with the data leakage when transmitting users' data, federated learning(FL) is regarded as a critical technology. Consequently, parameters rather than private data from similar local models are uploaded to the FL layer for further training.

\textit{2) Fog-Edge-Cloud Metaverse:} This distributed architecture is hierarchical and tiers different servers to eliminate the Metaverse fragmentation and computational bottlenecks.\textit{Kechadi et al.}\cite{kechadi2022edge} proposed this hybrid computational architecture for Metaverse services. In cloud layers, Cloud servers exist to simulate the virtual worlds. The edge layer servers are dedicated to processing specific virtual buildings' computation tasks and providing virtual environment animation. For fog servers located on the fog layer, servers are deployed to perform computation tasks of the virtual home environment and users' movements. 

\subsection{Improvement}
\textit{User Side:} In virtual worlds, accurate predictions and assessments of users' behaviours, habitual preferences, and movement paths are crucial to providing users with an immersive and real-time experience. Based on the attributes of MEC, stable and reliable seamless services could be offered to users' devices, and real-time services are also provided as a result of the location of MEC servers \cite{lee2021all}. \par
Furthermore, since edge servers have more computing capabilities than local devices(e.g. VR devices, HMDs, and AR glasses), they can efficiently process computation-intensive tasks and transfer feedback to mobile devices, which will reduce latency and thus alleviate user discomfort such as dizziness caused by latency.

\textit{System Side:} The two main merits of applying MEC to Metaverse systems today are enhanced resource utilisation and significantly reduced latency. Taking advantage of the fact that MEC servers come with their own computational resources\cite{ng2022unified}, most Metaverse applications can be processed through these servers. At the same time, the wearability and small size of most devices lead to limitations on the battery life and computational resources, which is tackled by the MEC-based Metaverse. Simultaneously, the dynamic resource allocation\cite{han2022dynamic} and edge intelligence-based technology\cite{van2022edge} further improve the performance of all aspects of the Metaverse substantially.

\begin{figure}[t]
    \centering
    \includegraphics[width=8cm, height=2cm]{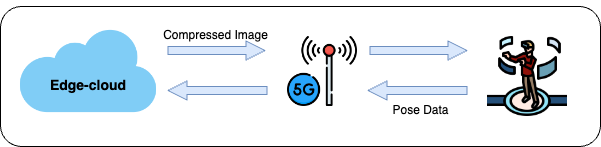}
    \caption{BoundlessXR consists of XR devices, 5G network, and edge-cloud.}
    \label{fig3}
\end{figure}

\subsection{Application Scenarios}
On-device processing is an essential part of performing Metaverse applications. In a stand-alone model, on-device processing is responsible for dealing with entire XR processing assignments. While augmented by the edge cloud, on-device processing provides rendering and tracking that are power-efficiency, high-performance and latency-sensitive. Nowadays, Boundless XR created by the Qualcomm uses split rendering to distribute the computation tasks between the edge servers and end devices to deliver a genuinely immersive XR experience over 5G. Specifically, the architecture is presented in Fig. ~\ref{fig3}. Firstly, VR devices send 6-degrees-of-freedom head pose data to the edge cloud over 5G. The edge cloud sends the compressed image to the devices after processing this data and rendering a new image. Finally, VR devices further perform on-device processing. Based on the 5G and MEC technologies, the motion-to-render-to-photon latency is less than 20ms, which meets Metaverse services' requirements. NVIDIA's CloudXR is also a novel solution for streaming XR data from any openVR/XR application on a distant server, which can also dynamically adjust network conditions and maximize image quality while minimizing effective latency.

In addition to innovative applications by commercial companies, deep research in edge computing is also evolving in academia. A remarkable FACT algorithm was proposed in the paper\cite{liu2018edge} to balance three main factors, including network latency, computation latency, and system accuracy in an edge-based mobile augmented reality system. Applying this new paradigm to the MAR applications significantly addressed the optimization problems as a result of multiple mobile users and improved the performance of processing video frames in Metaverse. In order to optimize the computation resources and AR configurations on the edge servers, another optimization protocol called DARE was presented in the paper\cite{liu2018dare}. When considering multiple coherence times in the research of MEC-based MAR, the algorithm based on Lyapunov optimization in the paper\cite{liu2019edge} can effectively improve the reliability and the quality of augments (QoA) in Metaverse systems. Furthermore, when studied from the perspective of energy consumption and the interactions between MAR system parameters, the LEAF algorithm was studied by \textit{Wang et al.}\cite{wang2020user} in the specific energy-aware MEC-based MAR system.

\section{FUTURE RESEARCH CHALLENGES}\label{sectionVI}
In this section, future challenges of MEC-based Metaverse are discussed from the following aspects.
\subsection{Computer Resource Allocation}
Currently, papers are primarily conducted on computation allocation. \textit{Tang et al.} \cite{tang2020computation} proposed an optimization algorithm by considering the computation offloading strategies and resource allocation problems together and effectively solved the interuser interference due to load. \textit{Nath et al.} \cite{nath2020deep} studied the cached content by four decision parameters were considered, including whether a given task needs to be cached, how much transmission power is appropriate to utilize during offloading, and how many MEC resources to assign to perform a task to further improve collaboration issues between MEC servers. However, these researches do not consider the dynamic nature of Metaverse systems. Specifically, the computational resources of each node are dynamic in real time rather than static. Therefore, idle computational resources will be unused inefficiently when the MEC is processing computationally intensive tasks after nodes are pre-determined. For further research, how to solve the problem of dynamic allocation of computational resources is a crucial challenge to be tackled.
\subsection{Mobility Management}
Metaverse applications rely heavily on users’ states and previous behaviours, meaning mobility is a key consideration when implementing applications sensitive to continuity and real-time \cite{huang2022mobility}. On the other hand, user mobility and excellent mobility management ensure seamless service to enhance the user experience and QoS further. The alternative approaches are caching\cite{cai2022joint} and SDN controllers now. For caching, the main challenges are how the popularity of the content to be cached can be accurately predicted and how we can achieve collaborative caching across multiple cells to improve web caching performance. \cite{liu2021rendering} For SDN controllers, \textit{Shah et al.} \cite{shah2022sdn} propose an SDN-based MEC-enabled 5G vehicular networks. They use four software modules developed by themselves in the SDN controller to develop the rules of QoS further. Furthermore, the SDN controller proposed can federate and coordinate the allocation of MEC resources to further provide continuous service and seamless coverage to mobile users. In future research directions, mobility management can be combined with artificial intelligence. Through this technologies convergence, the Metaverse system can learn the user’s behaviours on its own, predict how the user will move, and transmit information about the user’s state to the SDN controller in advance, thus providing a more reliable service.
\subsection{User Experience}
Digital avatars experience various activities in photorealistic sub-Metaverses and can obtain perceptions through sensor devices. The final evolution of the Metaverse is to integrate with the physical world, which means dark real-world forces also exist in virtual life. For example, robberies and car accidents in Metaverse badly influence humans in the real world and even cause physical pain. Moreover, the unfettered behaviour of avatars, like sexual assault, also causes psychological victimization of the user. Thus, establishing a code of ethics to constrain the behaviour of avatars is a crucial challenge to maintaining a safe community for users in virtual worlds\cite{yu20226g}. Furthermore, user addiction\cite{zhu2020activity} is another harmful influence on emotional health in Metaverse.
\subsection{Data}
For obstacles of Metaverse in the future, two main factors are supposed to be noticed:  data leakage and data reliability.\textit{ 1) data leakage:} MEC-based metaverse architecture is not the same as the decentralized bitcoin architecture. Private and safety-sensitive data (such as avatar ID numbers and passwords, digital assets, and currencies) are transmitted to the edge cloud from end devices, which may suffer grievous attacks by potential adversaries. \textit{Lee et al.}\cite{lee2021adcube} identify three main ad fraud threats, including blind-spot tracking, gaze and controller cursor-jacking, and the misuse of assisted display in content sharing. \textit{2) data quality:}The data quality levels directly alter the metaverse applications' QoS. High data quality significantly improves the QoE of users, while low level stands on the opposite side. \textit{3) data reliability:} Data sources and generation processes will cause the data to be unreliable if the sources are fake IDs and jitters occur during propagation. Hence, preventing leakage and improving the quality and reliability of data in Metaverse are also essential requirements during the development of realizing Metaverse.
\subsection{Delay}
The main goal of the Metaverse is to deliver the processed data to the XR device in real-time via the downlink and to render the virtual spatial environment. For example, the delay generated by VR devices should not exceed 20ms; otherwise, it will cause spatial vertigo and vomiting to the user in the Metaverse. Hence, the partitioning of computational tasks is a constant focus of researchers\cite{cai2022compute}. Furthermore, balancing the tradeoff between task splitting and delay is essential. For example, Lightweight and wearable XR devices can be worn easily but with fewer computational resources.
\subsection{Privacy}
Data privacy and user identity are the two significant aspects of Metaverse security that must be ensured. For data privacy, users' virtual assets, facial and retinal data, and transaction records in the Metaverse are at risk of leakage, especially during data processing and transmission. Moreover, after one user's behavioural data and interactions with other avatars have been collected and analyzed by hackers\cite{di2021metaverse}, a fake avatar will be made that can imitate the original avatar to carry out social activities in the Metaverse and even commit crimes. Bystander privacy in the Metaverse is also at risk of being acquired. Then, the protection of data and identity becomes the way forward.

\section{CONCLUSION}\label{sectionVII}
In this paper, we have introduced an in-depth survey of Metaverse based on mobile edge computing. We have presented an overview of Metaverse and discussed its fundamentals, characteristics, core technologies, and existing issues. Afterwards, fundamental research and application scenarios like AR(Web-AR)/VR/MR of mobile edge computing are presented. Significantly, we analyze the convergence of these two prominent technologies from aspects including architectures, improved peculiarities, and application scenarios. Furthermore, we show the future research challenges of mobile edge-based Metaverse. We anticipate that this survey will shed light on the direction of the MEC and Metaverse and spark even more impressive research in the future.

\section*{Acknowledgement}

This research is supported in part by Nanyang Technological University Startup Grant; in part by the Singapore Ministry of Education Academic Research Fund under Grant Tier 1 RG97/20, Grant Tier 1 RG24/20 and Grant Tier 2 MOE2019-T2-1-176.

\bibliographystyle{IEEEtran}
\bibliography{mybibfile}
\vspace{12pt}
\end{document}